\documentclass{article}
\usepackage[utf8]{inputenc}
\usepackage[a4paper,left=3cm,right=3cm,top=3cm,bottom=3cm]{geometry}
\usepackage{amsmath}
\usepackage{amssymb}
\usepackage{graphicx}
\usepackage{esint}
\usepackage{enumitem}
\usepackage{multirow}
\usepackage{graphicx}
\usepackage{hhline}
\usepackage{hyperref}
\usepackage{centernot}
\usepackage{cite}
\hypersetup{
  colorlinks   = true,    
  urlcolor     = blue,    
  linkcolor    = blue,    
  citecolor    = black     
}

\newcommand\ccbar{$c\bar{c}$ }

\usepackage{color}
\usepackage[makeroom]{cancel}
\usepackage[normalem]{ulem}

\title{Color Reconnection Effects in $J/\psi$ Hadroproduction}

\author{
Piotr Kotko$^a$,
Leszek Motyka$^b$,
Anna Staśto$^c$
\\ \\
$^a${\it AGH University of Krakow,}\\ 
{\it Faculty of Physics and Applied Computer Science,} \\ 
{\it al. Mickiewicza 30, 30-059 Kraków, Poland} \\ \\
$^b${\it Jagiellonian University, Institute of Theoretical Physics,}\\ 
{\it S. Łojasiewicza 11, 30-348 Kraków, Poland } \\ \\
$^c${\it The Pennsylvania State University, Physics Department,}\\ 
{\it 104 Davey Lab, University Park, PA 16802, USA }
}

\date{}

\begin{document}

\maketitle

\begin{abstract}
    We investigate production of $J/\psi$ mesons in hadron-hadron collisions, defined as low invariant mass \ccbar singlets produced in a mixture of
    perturbative and nonperturbative mechanisms provided by the PYTHIA Monte Carlo.
    We find that in this model the color reconnection mechanism, which breaks the factorization, is essential to reasonably describe the experimental data.
\end{abstract}

\section{Introduction}

The inclusive $J/\psi$ hadroproduction is a process of particular sensitivity to color flow in Quantum Chromodynamics (QCD). The straightforward approach to this process in QCD is based on the exact correspondence between the color representation of the partonic content of $J/\psi$ and the meson itself. This framework --- the color singlet model (CSM) --- fails badly in describing the measured cross sections, both in the overall magnitude and in the $p_T$ dependence \cite{Baier:1982zz,Baier:1983va,Glover:1987az,CDF:1992cmg,CDF:1997uzj}. The color singlet model underestimates significantly the data and gives a much too steep decrease with $p_T$. At the most general level 
 the natural conclusion is 
that some of the partonic states \ccbar of the charmed quarks produced in the color octet representation may form a color singlet state --- a meson. The question how it actually occurs that a partonic state with different color representation may form a color singlet hadron has not been definitely answered from the first principles. 
Currently, the most successful approach in describing the inclusive charmonium hadroproduction data is the color octet model (COM) \cite{Bodwin:1994jh,Cho:1995vh,Cho:1995ce}, evaluated at the next-to-leading accuracy, see e.g.\ Refs.\ \cite{Butenschoen:2009zy,Butenschoen:2010rq,Chao:2012iv,Ma:2018qvc}. This model originates from an effective field theory for Quantum Chromodynamics 
with heavy quarks --- non-relativistic QCD (NRQCD) \cite{Caswell:1985ui,Lepage:1992tx ,Bodwin:1994jh}. It introduces non-zero amplitudes of transitions between the partonic $c\bar c$ states and the meson states. In these transitions color and spin quantum numbers may change. Within NRQCD this mismatch is understood as a result of additional contributions to color and spin coming from the gluonic degrees of freedom in the hadron  \cite{Bodwin:1994jh} or additional color exchanges and emissions with long wavelengths \cite{Cho:1995ce,Cho:1995vh}.
Hence, in this model one makes the simplest assumption, that the $c\bar c \to$~quarkonium amplitudes are universal and do not depend on the environment in which the parton--hadron transition occurs. Thus, the COM may be also viewed as a fragmentation process of $c\bar c$ partonic states into a charmonium, that is factorized from other initial and final state processes including hadronic remnant dynamics \cite{Nayak:2005rt}. The phenomenological success of this approach in describing the single particle inclusive distributions of charmonia is a good justification of these assumptions. In the primary formulation COM required a relatively large number of additional parameters that describe the Long Distance Matrix Elements corresponding to transition amplitudes from various partonic states to different quarkonia states, but recently within NRQCD these matrix elements were expressed via universal gluon condensates, which greatly enhanced the predictive power of the model \cite{Brambilla:2022ayc}.

Along a similar line, the heavy quarkonia hadroproduction can be also described within a simpler Color Evaporation Model (CEM) \cite{Fritzsch:1977ay,Halzen:1977rs, Gluck:1977zm, Barger:1979js,  Amundson:1995em}, that also assumes a local transition between color octet $c\bar c$ pair and a charmonium state. For reviews of all the described quarkonia production mechanism see \cite{Lansberg:2006dh,Brambilla:2010cs,Lansberg:2019adr}. The abovementioned factorization (independence of the environment) of the parton--hadron transition is, however, challenged by measurements of inclusive hadroproduction of charmonia pairs, in particular the $J/\psi$ pair production.  

The measurements of double heavy quarkonia hadroproduction were performed at the Tevatron \cite{D0:2014vql,D0:2015dyx} and at the LHC \cite{CMS:2016liw,ATLAS:2016ydt,LHCb:2016wuo}. The data were interpreted   within the factorized double parton scattering formulation. In this framework, the probability of the double parton scattering was found to be significantly larger than in all other processes. To be more specific,  it is customary to express the double scattering cross section $\sigma_{AB}$ as a product of single scattering cross sections $\sigma_A$ and $\sigma_B$ divided by  the double scattering parameter $\sigma_{\rm eff}$,
\[
\sigma_{AB} = s_{AB}\frac{\sigma_A \sigma_B}{\sigma_{\rm eff}}\,,
\]
where the symmetry factor $s_{AB} = 1/2$ for $A=B$ and  $s_{AB} = 1$ otherwise.
In the measurements of the double quarkonia production  \cite{D0:2014vql,D0:2015dyx,CMS:2016liw,ATLAS:2016ydt,LHCb:2016wuo} $\sigma_{\rm eff}$ was determined to be in the range between 2~mb and 11~mb, while in other processes with double parton scattering the typical value is 15--20~mb, for a comparison see e.g.\ Fig.~14 in \cite{ATLAS:2016ydt}. Recently, also triple $J/\psi$ production was observed by the CMS collaboration \cite{CMS:2021qsn}, and the extracted value of $\sigma_{\rm eff}=2.7^{+1.4}_{-1.0} \,{} ^{+1.5} _{-1.0}$~mb is again small. This discrepancy strongly indicates that in the case of double and triple quarkonia production the assumption about independent production does not work. There might be different reasons for this. One of possibilities is a dependence of the transition amplitudes from the color octet $c\bar c$ states to $J/\psi$ on the environment given by the complete event, e.g.\ by interactions of the $c\bar c$ pair with the proton remnants. A certain hint that this could take place comes from some tension between the COM parameters in $J/\psi$ production in $pp$ or $p\bar p$ collisions and in $e^+e^-$ collisions reported in Refs.\ \cite{Zhang:2009ym,Ma:2010yw,Ma:2017xno}. The leading COM coefficient for $J/\psi$ production at LEP was constrained to be significantly smaller than in $J/\psi$ hadroproduction. Note however, that in the analysis of Ref.\ \cite{Butenschoen:2011yh} this tension is seen as less pronounced. 

Within the perturbative approach, that describes the short-distance part of the production process, long range correlations in rapidity in the $J/\psi$ pair production may be generated by perturbative multiple scattering effects, mostly exchanges of gluons in the $t$-channel. These effects were investigated for single inclusive $J/\psi$ hadroproduction \cite{Khoze:2004eu, Motyka:2015kta, Kotko:2019kma} and it was found that without the COM contributions the perturbative rescattering effects are not sufficient to describe the data. 
Additional gluon exchanges in $J/\psi$ pair hadroproduction were also considered \cite{Baranov:2012re,Szczurek:2017qul} and were found not to explain the data in the color singlet approach.

Hence, in this paper we perform a study of another microscopic mechanism of color redistribution in $J/\psi$ production that is not necessarily factorized. The approach which we take follows pioneering studies of quarkonia hadroproduction due to possible color reconnection by Edin, Ingelman and Rathsman \cite{Edin:1997zb} within Soft Color Interaction model \cite{Edin:1995gi}, see also \cite{Enberg:2001vq}. The color reconnection effects were earlier studied for hadronic $W^+ W^-$ events in $e^+ e^-$ collisions ~\cite{Sjostrand:1993hi} and for rapidity gaps at HERA \cite{Lonnblad:1995vr}, for more recent studies see Refs.~\cite{Argyropoulos:2014zoa,Christiansen:2015yca}. 
Also $J/\psi$ production in proton--proton collisions at the LHC was studied using PYTHIA with effects of color reconnection included in the cluster collapse contribution \cite{Weber:2018ddv}. 
The color reconnection mechanism  is assumed to occur before or at the transition between the partonic and hadronic phase of the scattering event.  The color redistribution in hadronic scattering may have different sources. It can emerge from secondary scatterings, exchange of semihard gluons or it could be a nonperturbative phenomenon, related to color tunneling or a string reconnection in the Lund string picture. As currently there is no complete fundamental understanding of color redistribution between the partonic and hadronic phases, it is necessary to model this phenomenon and confront the model predictions with the data.

The framework for the current study is provided by PYTHIA Monte Carlo event simulator \cite{Sjostrand:2014zea}. The crucial ingredients of the applied model for single inclusive $J/\psi$ production are:
\begin{enumerate}
\item Generation of proton-proton collision
 events in which $c\bar c$ pairs are created. In general, the charm (anti-)quarks may be created in the primary hard scattering, subsequent scatterings, parton showers interleaved with the multiple scatterings, or, finally, by the beam remnant treatment.

\item Including color reconnection effects according to PYTHIA QCD-based model \cite{Christiansen:2015yqa}, labeled as CR1. These effects reconnect the ends of Lund strings and may turn some of the $c\bar  c$ pairs from a color octet state to a color singlet state. 

\item Imposing conditions for $J/\psi$ formation from a $c\bar c$ pair: the color singlet constraint and the requirement of the invariant mass $M_{c\bar c}$ of $c\bar c$ pair to be close to the $J/\psi$ mass $M_{J/\psi}$. The value of the invariant $c\bar c$ mass range $\Delta M_{c\bar c}$ that allows the $J/\psi$ formation is taken as the model parameter. A natural upper limit $\Delta M_{c\bar c}< 1$~GeV is imposed by the open charm production threshold. In fact one expects $\Delta M_{c\bar c}$ to fall below 0.5~GeV that corresponds to the energy gap between the excited and ground states in the charmonium system. 

\end{enumerate}

We stress that in our approach the COM and CSM mechanisms are not explicitly included in generation of $J/\psi$ mesons. 
The key difference between the present approach and those models comes from the fact that the present picture is more microscopic and ``non-local''. In contrast to COM we include the parton shower as a source of $c$ or $\bar c$ quarks that may eventually recombine into the heavy quarkonia. Besides that, the probability of color reconnection and, in consequence, forming $c\bar c$ color singlets, depends heavily on the parton activity and total color flow in the whole event. Clearly, this implies a strong environment dependence of quarkonia production. This, in turn, could result with a long range rapidity correlations in pair production of heavy quarkonia caused by an influence of the same environment: the partons and the color flow in a given event.

Let us stress that the $J/\psi$ production model used in the present work is rather basic. We use the default settings of PYTHIA, without extra tuning; only the color reconnection model is set to the QCD-based model, which is not the default setting.
We introduce only one additional free parameter, $\Delta M_{c\bar{c}}$. The value of $\Delta M_{c\bar{c}}$ that is necessary for a good description of the data is found to be 0.3~GeV, well in the expected range. The obtained overall description of the $J/\psi$ hadroproduction data is rather good already with this simplest model. Both the magnitude and the shape of the differential cross section $d\sigma/ dp_T$ are very close to the data. Without the color reconnection, the cross sections are too small by one order of magnitude. It is remarkable, that the ``hardness'' of $d\sigma/ dp_T$ at large $p_T$ is naturally obtained within this approach, a feature that is out of reach in the CSM.
One expects that with dedicated tunning of the color reconnection model the description of the single inclusive charmonia hadroproduction could be further improved. Conversely, one may use the process to constrain the color reconnection mechanism. Finally, possible correlations of charmonia formation in this model may lead to an improved description of hadroproduction of quarkonia pairs. In this paper, however, we focus on the single inclusive $J/\psi$ production only. This is because we need to establish the details of the approach and understand well the microscopic mechanisms included in the applied scheme before addressing more subtle process of double quarkonia production, also keeping in mind that the latter process requires much higher statistics and numerical resources.  Hence the double $J/\psi$ production is left  for a future dedicated analysis.

\section{Color reconnection}
\label{sec:color_reconnection}

In this section we shall briefly overview the color reconnection (CR) model used in our computations with PYTHIA, focusing on aspects that are essential to understand the observed effect on the $J/\psi$ production cross section. For a comprehensive description of the model see \cite{Christiansen:2015yqa}.

The main problem addressed by color reconnection models is to determine a  realistic, consistent with QCD, color topology of the whole event, including the nonperturbative components of the calculation. The main complication arises due to the Multiple Parton Interactions (MPI). A naive expectation, that the color flow lines of subsequent hard processes are disconnected and always hadronize separately is in disagreement with the experimental data. It is important to stress, that the concept of the MPI in Monte Carlo event generators is quite different than for e.g.\ the hard double parton scattering, where for sufficiently inclusive observables the QCD factorization holds (see \cite{Diehl:2011tt,Diehl:2011yj,Blok:2013bpa,Diehl:2014vaa,Buffing:2017mqm,Diehl:2021wvd} and an overview in \cite{Diehl:2017wew}). The soft MPI modeled in event generators are nonperturbative phenomena, thus existence of rather strong color correlations between different subsystems is a natural expectation.
Obviously, the color correlations between MPI subsystems occur also perturbatively, see e.g.\ \cite{Blok:2022mtv}, but they seem not strong enough.
Thus, although basing on the perturbative QCD one can determine the possible color flows for the hard part of the event, including the showers, there is still a large degree of ambiguity. For example, a color flow line from one MPI subsystem, instead of a beam remnant, can be connected to another MPI subsystem; there is no perturbative way of telling to which subsystem, however. 
In PYTHIA event generator, this ambiguity is resolved by the hadronization model. The possible color reconnections are those that minimize the measure given by 
\begin{equation}
    \lambda= \sum_i \ln[(p_i+p_{i+1})^2/m_0^2] \,,
\end{equation}
where the sum goes over all partons that constitute a "string" ($m_0$ being a hadron mass scale), i.e.\ an object that next fragments into hadrons, in the spirit of the Lund model \cite{Andersson:1983ia}.

In MC event generators the color quantum numbers are technically implemented in terms of color tags, which are not bounded from above, in the spirit of the leading color approximation. Quarks and anti-quarks carry a single tag, for either color or anti-color, while gluons carry a pair of tags. The planar topologies given by the lines with specified color or anti-color tags provide the leading color flow diagrams. 
In the new QCD-based CR model in PYTHIA, the simplified SU(3) QCD rules are used to determine the possible color topologies. Whether a non-leading color topology is realized is determined by the string length rule -- preferred topologies are those that decrease the $\lambda$ measure.

As the CR model is driven by the minimization of the string length, it should often color-reconnect partons that -- if not for color-anti-color mismatch -- could form a hadron. Such partons are more likely to originate in a single mother. As we shall see in the next Section, this is exactly what drives the production of low invariant mass \ccbar singlets.

\section{Results}
\label{sec:Results}

We have performed our computations using PYTHIA version 8.307 with the default tune and most parameters set to their default values. As already mentioned, the only exception is that we use the QCD-based color reconnection model in our study, which is not a default setting at present. We stop the event generation just before the hadronization stage, either with, or without the color reconnection mechanism switched on. We shall define the $J/\psi$ state as the \ccbar color singlet state with the invariant mass within the range $3.0<M<3.3$~GeV. We will discuss the invariant mass window dependence later in this Section. The color singlet is defined here as a pair of partons with matching color and anti-color tags.

Before we focus on the $J/\psi$ cross section, let us first discuss the possible sources of \ccbar pairs in the color singlet state. The core hard process in PYTHIA is provided by the leading order $2\rightarrow 2$ partonic process, which, obviously, cannot generate a \ccbar singlet. Perturbative corrections are provided by the parton showers. These give both the collinear logarithm resummation, as well as part of the NLO correction. This second mechanism of the \ccbar production can in principle produce singlet states, however, the c-quarks or anti-quarks will originate from different branchings, thus are rather unlikely to form a state with a low invariant mass. The next key ingredient of PYTHIA is the MPI treatment, which is interleaved with the showers. PYTHIA also takes care of the beam remnant, so that the quantum numbers, including the momentum of the partons, are preserved.
As mentioned in the preceding section, when dealing with the color topologies, there are ambiguities that are resolved by the CR mechanism.
The following exercise is useful to better understand how the CR affects the production of the \ccbar singlets within the different mechanisms.

\begin{figure}
    \centering
    \includegraphics[width=9cm]{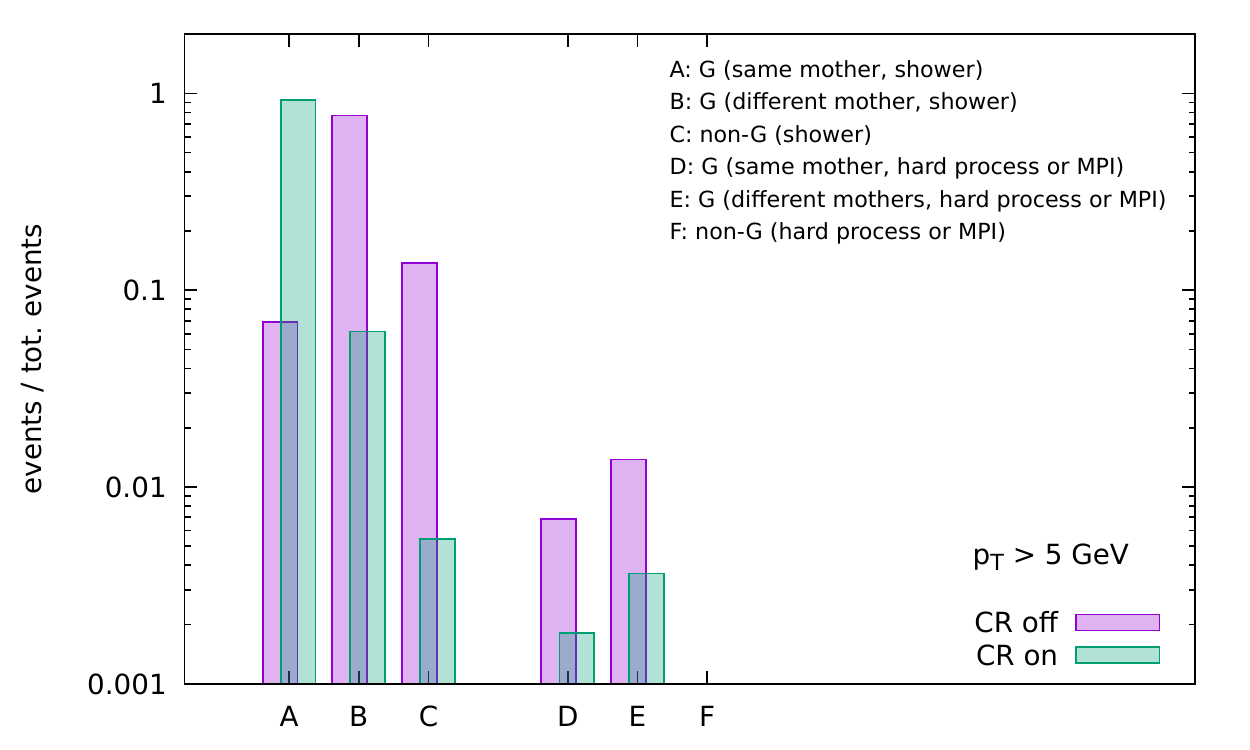}\\
    \includegraphics[width=9cm]{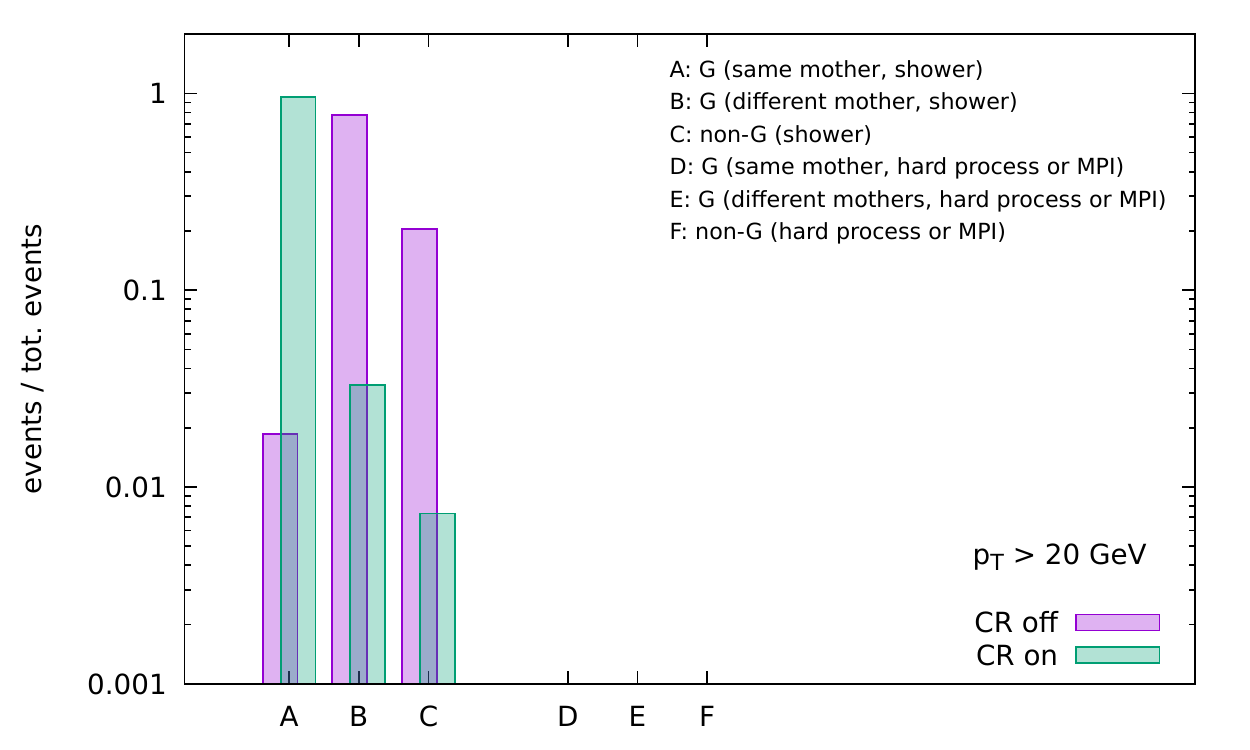}\\
    \includegraphics[width=9cm]{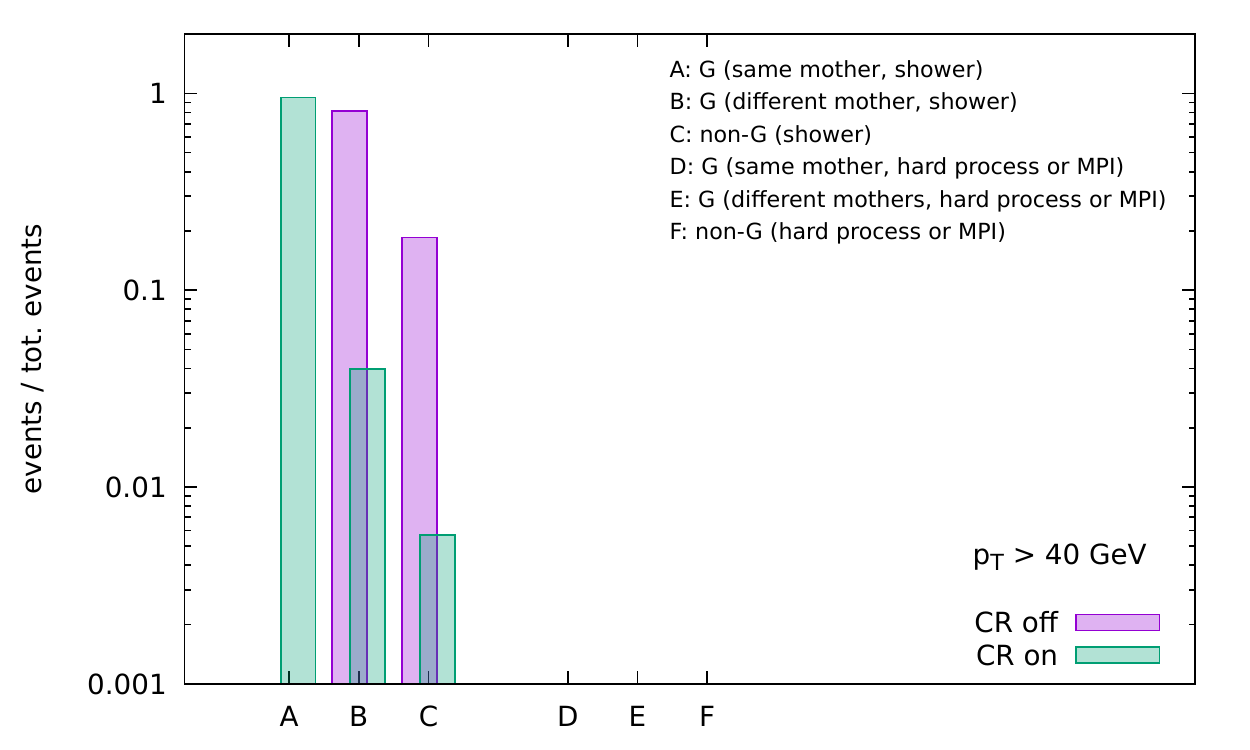}
    \caption{
    \small
    Study of the origin of the low invariant mass singlet \ccbar pairs in PYTHIA. The histogram shows the event number for different classes of processes, normalized to the total number of events (CR on and off are normalized independently). The classes A,B and D,E correspond to \ccbar singlet coming from a gluon. We distinguish then whether the singlets come from a single mother or different mothers (class A and B respectively), through the showers, or directly through the hard process or subsequent hard processes (class D for a single mother and E for different mothers). Classes C and F correspond to a non-gluon origin in either the shower or hard processes. The three plots correspond to different $p_T$ cuts on the transverse momentum of the singlet, starting from $p_T>5$~GeV (top), $p_T>20$~GeV (middle) and $p_T>40$~GeV (bottom). Notice, that the vertical scale is logarithmic. The actual cross section for the color reconnection (CR) turned on scenario is much higher then the CR off result; the plot shows only relative contributions in each scenario. 
    \label{fig:c_origin}
    }
\end{figure}

\begin{figure}
    \centering
    \includegraphics[width=16cm]{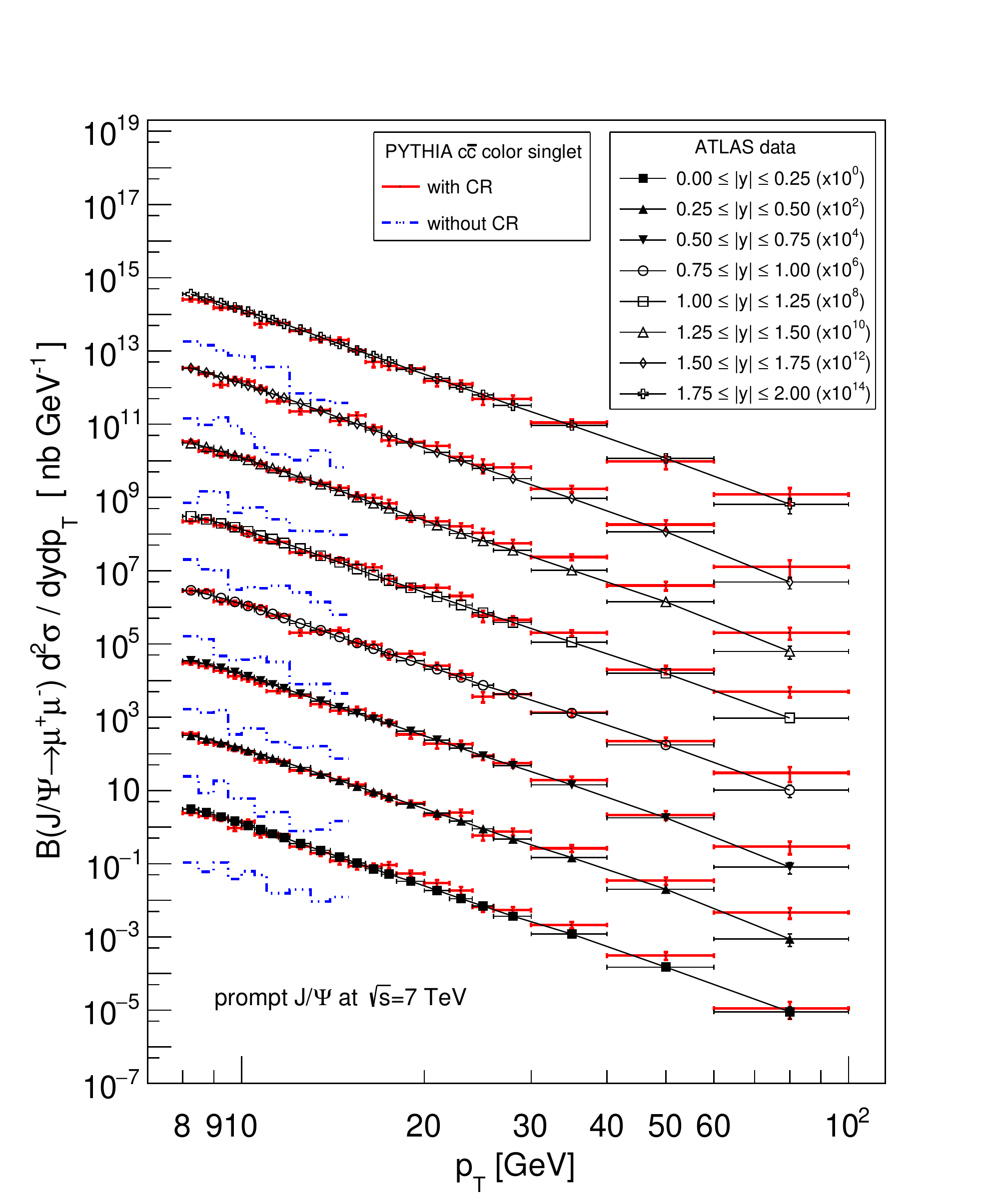}
    \caption{Transverse momentum spectra of $J/\psi$ obtained from the PYTHIA generator, where the $J/\psi$ state is constructed as the \ccbar color singlet with invariant mass $3.0<M<3.3$~GeV, in comparison with the ATLAS data \cite{ATLAS:2015zdw}. The dot-dashed histograms correspond to the scenario without the color reconnection model, which significantly underestimates the cross section (due to the low statistics, we show only the low $p_T$ range).}
    \label{fig:singlet_vs_ATLAS}
\end{figure}

In Fig.~\ref{fig:c_origin} we show the relative contributions of different mechanisms that generated a singlet, with, and without the color reconnection for different $p_T$ cuts on the color singlet state. In the latter case, the low invariant mass singlet is predominantly produced through shower (and not the hard process) from gluons of different origin (i.e.\ different mothers --- see the class B in the figure). Naturally, production of singlets that way is combinatorically rather unlikely, what reflects itself in a rather small cross sections for a production of \ccbar singlets (cf. Fig.~\ref{fig:singlet_vs_ATLAS} which we shall discuss in more details further below). 
 It is interesting to observe, that even with the CR off, the class A, i.e.\ same mother gluon origin, is non-empty. It is not possible in the pure leading color approximation, meaning that these events were created due to the color reconnection provided by the MPI alone. This is confirmed by the observation that these events are suppressed when increasing the $p_T$, being completely eliminated above $p_T\gtrsim 40$~GeV.
On the other hand, when the color reconnection mechanism is turned on, the dominant production of singlets is via shower from the same mother, which most of the time is a gluon. Such configurations are apparently enhanced by the smallness of the string $\lambda$-measure.
This mechanism seems somewhat similar in spirit to the COM, but one should remember that the enhancement of color configurations leading to the low invariant mass \ccbar  singlets is possible mainly thanks to the MPI mechanism. This non-perturbative mechanism is not explicitly present in COM.

\begin{figure}
    \centering
    \!\!\!\!\!\!\!\!\!\!\includegraphics[width=17cm]{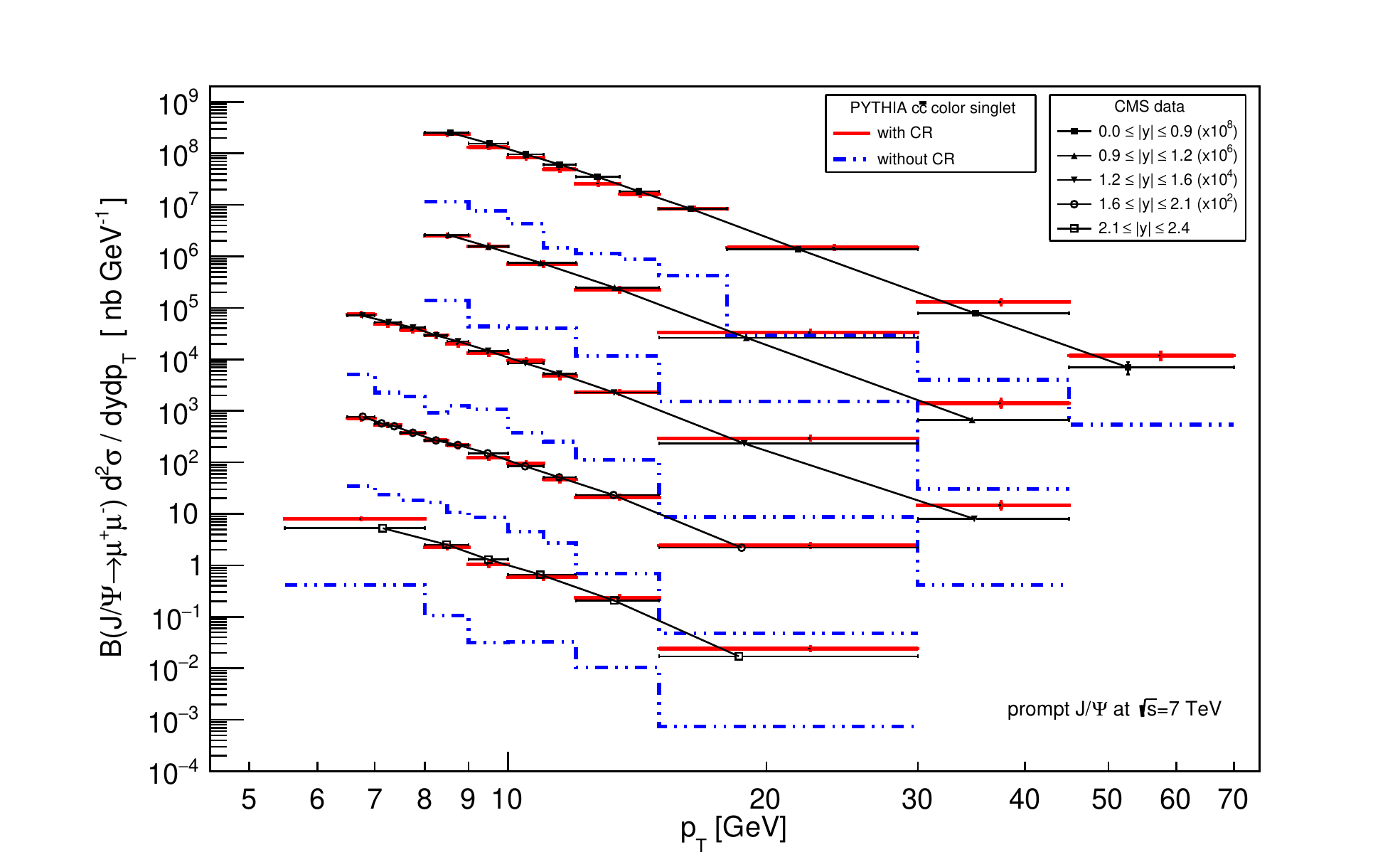}
    \caption{Same as in Fig.~\ref{fig:singlet_vs_ATLAS}, but in comparison with CMS data \cite{CMS:2011rxs}.}
    \label{fig:singlet_vs_CMS}
\end{figure}

Let us now turn to our main results for the transverse momentum spectra of the $J/\psi$ at $\sqrt{7}$~TeV. As the efficiency of the \ccbar singlets generation in PYTHIA is overall rather low, we have used cluster computing to obtain a reasonable statistics in rapidity bins of the interval $0<|y|<2.4$, as defined by the ATLAS collaboration in their measurement of the prompt $J/\psi$ \cite{ATLAS:2015zdw}. The $c\bar c$ parton pair invariant mass $M$ cut for the $J/\psi$ formation was tuned to obtain good overall normalization of the cross-sections. The final choice of $3.0<M<3.3$~GeV is fully consistent with the energy spacing in charmonium system. In Fig.~\ref{fig:singlet_vs_ATLAS} we compare the data to our calculations with, and without, the color reconnection. It is clear, that the color reconnection model is essential to get the magnitude of the cross section right. Overall,  we obtain a rather good description of the ATLAS data in all rapidity bins, especially at lower $p_T$. 
We also compute the differential cross sections for the kinematics of the older CMS data \cite{CMS:2011rxs} (Fig.~\ref{fig:singlet_vs_CMS}). 
Here we find a similar agreement, although for a very large rapidities $|y|>2.0$, the model seems to deviate more from the data.
We stress that the PYTHIA parameters were not tuned or modified in any way; the only parameter is the invariant mass of the singlet, which however lies within the reasonable range for the $J/\psi$. Likely, tuning the parameters would improve the description even further.

To conclude this Section, let us stress that the production of $J/\psi$ in PYTHIA (defined as the \ccbar singlet with the invariant mass cut) is provided by a highly nontrivial mechanisms. First, there is a perturbative mechanism, which is a hard process followed by the parton showers (although this  might be arguable, we regard the showers as perturbative mechanisms here). 
As is known, this mechanism itself, however, does not provide a realistic color flow distribution in hadron-hadron collision. Thus, the subsequent hard processes are needed, as well as the realistic way of picking up enhanced color topologies --- this is provided by the color reconnection mechanism tied together with the color string measure prior to hadronization. 

Naturally, the above picture should be considered as a model, specific to PYTHIA event generator. However, a similar mechanisms are implemented in other generators, including the Herwig program \cite{Bahr:2008pv,Gieseke:2012ft,Gieseke:2018gff}. Within this model, formation of the final $J/\psi$ state depends on the whole event color topology and kinematics, including the beam remnants and parton showers. Taken at face value, this introduces possible non-factorizable effects in the single $J/\psi$ production. This picture is supported by a good description of data obtained within this model. It is an interesting and open problem how effects at the onset of non-perturbative phase, for instance the hadronization effects, affect the proofs of factorization for fragmentation functions \cite{Collins:2016ztc}. We believe that further tests of the model proposed in the present work with double $J/\psi$ hadroproduction may shed more light on this problem.

\section{Summary and outlook}
\label{sec:Summary}

We have analyzed the production of the low invariant mass \ccbar singlets in PYTHIA, focusing on  the impact of the color reconnection mechanism.
Treating the \ccbar singlet with a proper invariant mass cut
3.0~GeV~$ < M_{c \bar c} < 3.3$~GeV as the $J/\psi$, we found a very good agreement with the experimental data for the transverse momentum spectra for single inclusive $J/\psi$ production, \emph{provided the color reconnection mechanism is turned on.} Both the normalization and the shape in the transverse momentum are very well reproduced. 
We found that a similar computation without color reconnection fails drastically in describing the data.
We have analyzed the origins of the $c\bar{c}$ singlets in the two scenarios. In the presence of the
color reconnection mechanism, the dominant production of singlets is from the same mother parton, which is a gluon, in a shower or a beam remnant. On the other hand, in the case when the color reconnection is switched off, the dominant origin for the $c\bar{c}$ singlets are different mothers. 
Such parton configurations are unlikely  to create the low invariant mass singlet states; they are created somewhat accidentally as there is no mechanism that would give them any  preference.

Let us point out that the emerging physics picture of the $J/\psi$ formation at larger $p_T$ exhibits certain similarity to that in the color octet model and in the color evaporation model, where the dominant production mechanism at high $p_T$ is a gluon to $J/\psi$ transition. There is, however, one main difference: in the present model this transition may depend on the environment: the beam remnants and other partons in the event, while in the other models this there is no such dependence. This difference is related to the very question of universality of fragmentation \cite{Collins:2016ztc}, in our case of the gluon or a color octet \ccbar pair to $J/\psi$. As we already discussed in the Introduction, the double $J/\psi$ production data suggest that there occur strong long distance rapidity correlations that seem to contradict the picture of independent production. The model presented in this work could, in principle, generate some correlations of this kind.

The role of the environmental effects could be studied in more detail. For example, one could ask how the $c\bar c$ color octet to singlet transition rate depends on the beams used in the experiment, or on the associated production of other particles. One could also study how the double and triple $J/\psi$ production is affected by the color reconnection mechanism. The latter analysis would be particularly interesting in light of the recent experimental data from LHC. We  note however that,
the double $J/\psi$ production is at present a rather challenging computation in PYTHIA. Currently one would need a rather large scale cluster computing to produce  a reasonable statistics to make any meaningful conclusions. Investigation of that and other possibilities of enhancing the statistics is left for future study.

In this paper we did not study the polarization of $J/\psi$. In principle the model may be extended to obtain predictions for the meson polarization as well. A natural and simple extension that could be proposed, would follow the ideas presented in \cite{Martin:1996bp, Martin:1997sh}. In these references the concept of parton-hadron duality is used to associate the $\rho$ meson polarization produced in diffractive $\gamma^*$-proton scattering with the total angular momentum of the light quark-antiquark pair produced in partonic scattering. A similar approach could be applied to the $J/\psi$ hadroproduction within our model. For that, however, it is necessary to have an access to the charmed quark spins in the event simulation and to make additional assumptions on spin effects of color reconnection. In the simplest scenario one would assume that the color reconnection does not change the spins of partons. Unfortunately, at present, the polarization information on final state partons in PYTHIA is not available. To our knowledge, the situation regarding spin is better in Herwig Monte Carlo \cite{Richardson:2018pvo}. As a matter of fact, in future, we would like to extend our study of $J/\psi$ production in Herwig, and then also the polarization can be considered.

\section{Acknowledgements}
\label{sec:acknowledgements}
We thank  Christian Bierlich, Nora Brambilla and Tomasz Stebel for helpful comments. PK is supported by the Polish National Science Centre, grant no.\ DEC-2020/39/O/ST2/03011. LM gratefully acknowledges support of the Polish National Science Centre (NCN) grant no.\ 2017/27/B/ST2/02755.
We gratefully acknowledge Polish high-performance computing infrastructure PLGrid (HPC Centers: ACK Cyfronet AGH) for providing computer facilities and support within computational grant no. PLG/2021/014605. AMS is supported  by the U.S. Department of Energy Grant DE-SC-0002145 and within the framework of the Saturated Glue (SURGE) Topical Theory Collaboration.

\bibliographystyle{JHEP} 
\bibliography{refs}

\end{document}